\documentclass[preprint,           
               showpacs,            
               preprintnumbers,     
               aps,                 
               prd,          	    
               letterpaper,             
               nofootinbib,         
               tightenlines,        
               floats,floatfix,      
               showkeys]{revtex4-1}
\usepackage{graphicx}
\usepackage{dcolumn}
\usepackage{bm}
\usepackage{amsmath}
\usepackage{subfigure}
\DeclareGraphicsExtensions{.png,.jpg,.pdf}
\usepackage{amssymb}
\begin{document}

\preprint{}

\title{Cosmological restrictions for two DGP branes in a bulk with compactified extra dimension}

\author{Jairo Capilla-Valdepe\~na}
\email{jairogcv@esfm.ipn.mx}
\affiliation{Departamento de F\'isica, Escuela Superior de F\'isica y Matem\'aticas del Instituto Polit\'ecnico Nacional, Unidad Adolfo L\'opez Mateos, Edificio 9, 07738, Ciudad de M\'exico, M\'exico}

\author{Rub\'en Cordero}
\email{rcorderoe@ipn.mx}
\affiliation{Departamento de F\'isica, Escuela Superior de F\'isica y Matem\'aticas del Instituto Polit\'ecnico Nacional, Unidad Adolfo L\'opez Mateos, Edificio 9, 07738, Ciudad de M\'exico, M\'exico}

\author{Felipe Maya-Ord\'o\~nez}
\email{felipe-maya@my.unitec.edu.mx}
\affiliation{UNITEC, Calzada L\'azaro Cardenas 405, Lomas de Tlaquepaque, 45559, San Pedro Tlaquepaque, Jalisco, M\'exico}

\date{\today}

\begin{abstract}
We investigate the cosmological consequences of a two brane system embedded in a higher dimensional background spacetime with a compactified extra dimension described with an action that include the intrinsic curvature for each brane. We find that the dynamics of each brane is related to the other by means of cosmological restrictions that involve the scale factors, the equation of state parameters and the Hubble parameters of the two branes. We analyze the evolution of the scale factor and the equation of state parameter for the hidden brane when in the visible brane is present a cosmological constant, radiation or non-relativistic matter. The topological restrictions give rise to the cosmological scenarios without initial singularity, recollapsing and oscillating universes.
\end{abstract}


\maketitle

\section{Introduction}

Regge and Teiltemboim proposed long time ago \cite{RT} that our Universe can be described as a (3+1) dimensional hypersurface  embedded in a higher dimensional spacetime. In the same direction, a short time later, Rubakov proposed \cite{Rubakov:1983bb} that the Universe could be a  topological defect which emerged from a higher dimensional field theory. However these ideas did not attract interest due to the absence of phenomenological motivation. A renew interest of this kind of proposals emerged in the brane world scenario \cite{Rubakov:2001kp,Maartens:2010ar} where  one of its main characteristics is that the matter fields are trapped on a hypersurface (brane) immersed in a higher dimensional spacetime (bulk) where only gravity can be present in the brane and the extra dimensions. In fact, these sort of ideas were motivated in M-theory which is a generalization of string theory \cite{Schwarz:2008kd,cuerdas1,cuerdas2,cuerdas3} where our universe could be a very large 3-brane and the material fields made of open strings are attached to the branes with the exception of gravity which could propagate to the bulk. In the Horava-Witten model \cite{Horava:1995qa} there is an extra dimension compactified on an $S_1/Z_2$ orbifold and the matter fields are confined in two 10-dimensional hypersurfaces (9-branes) which are the boundaries of the 11-dimensional background spacetime. Additionally, 6 dimensions can be compactified on a Calabi-Yau manifold with size smaller than the distance between the two boundary branes which allow to consider as an adequate approximation a 5-dimensional background spacetime.  Specific brane world models constructed from M-theory were developed in \cite{Lukas:1998yy,Lukas:1998qs,Arnowitt}. Furthermore, the brane world models have attracted a lot of interest since they provide new possibilities to describe the dynamics of dark matter and dark energy \cite{Maartens:2010ar,Dutta:2016dnt}. The first brane world scenario was proposed in \cite{ADD-1} which has the main goal to resolve the hierarchy problem, i.e. to understand the reason why the electroweak scale is very small compared with the Planck scale. The model consisted of two branes with finite extra dimensions. Gravity is higher dimensional on short distances and transforms to 4-dimensional at scales which depends of the number of extra dimensions. Randall and Sundrum introduced a model which was motivated by the Horava-Witten model and it was constructed with two branes embedded in a non-separable bulk spacetime with a compactified extra dimension \cite{randall}. In this model gravity is mainly confined on the branes because the spacetime is warped around them. The size of the extra dimension decreased exponentially. Short time later another model was proposed by them with only one brane and a infinite extra dimension \cite{randall2}. Binetruy et. al. \cite{Binetruy:1999hy,Binetruy:1999ut} solved
the 5-dimensional Einstein equations with two branes and a compactified dimension in a global form. In that work they found a set of topological restrictions on the type of energy matter on the branes that are compatible with the underlying topology of the bulk. Recently, in \cite{Perez1,Perez2,Perez:2014mta,Perez:2012vx} were found new cosmological solutions for that system and the limits for low and high energies. Besides, it was analyzed the dynamics of the two branes for several matter-energy configurations in this setup.

Dvali, Gabadadze and Porrati (DGP) proposed \cite{DGP,DGP2} a different cosmological brane model by adding an intrinsic curvature term in the action describing the dynamics of the brane. This term can arise from quantum corrections and it produces modifications of gravity at large scale where it has a 5-dimensional behavior. The DGP cosmological model has two branches: the normal and the self-accelerated. The self-accelerating branch is able to produce a late time acceleration of the universe without the need of a cosmological constant or any material field. Then, the dark energy behavior could be only a geometrical and extra dimensional effect. This characteristic attracted a lot of interest, unfortunately the cosmological tests have strongly disfavored the physical viability of this branch \cite{Fang:2008kc}. However, the introduction of a cosmological constant on both branches restore the viability of these models at the cosmological level \cite{Lue:2004za,Lazkoz:2006gp,Lombriser:2009xg,Azizi:2011ys,Xu:2013ega} but still the self-accelerating branch has severe problems due to the existence of a ghost mode in the scalar sector of the gravitational field \cite{Maartens:2010ar}. Moreover, there exist solutions in the DGP model without $Z_2$ where the cosmological dynamics is identical to the standard cosmology \cite{Cordero:2001qd}. Other very interesting way to study the physical consequences of the DGP model is the process of galaxy formation which it was carried out in \cite{Hernandez-Aguayo:2020kgq} for the normal branch where there are important differences with general relativity. 

On the other hand, the introduction of an additional DGP brane can give rise to novel physical consequences like the modification of fine structure constant \cite{Dick:2015sla}. In \cite{Warkentin:2019caf}, it was calculated the changes of the laws of gravity of the DGP model by the presence of a second DGP brane. It turns out very interesting to study the global physical characteristics of two DGP branes immersed in a 5-dimensional background with a dimension compactified on an $S_1/Z_2$ orbifold and perfect fluids on each brane, and investigate the cosmological implications of the corresponding topological restrictions on the two brane setup. 

In this paper we study the cosmological evolution of two DGP branes embedded in a 5-dimensional background spacetime with one spatial compactified dimension. We find new cosmological solutions by means of an ansatz involving the scale factor and calculate the corresponding generalized Friedmann equations for both branes. We compute the cosmological restrictions for the matter-energy on each brane which are compatible with the topology of the background spacetime. We study the dynamics of the two branes by analyzing the behavior of their scale factors, the Hubble parameter, the equation of state parameter and the energy density of the hidden brane when we consider non-relativistic matter, cosmological constant and radiation for the visible brane.

The paper is organized as follows. In Section 2 we present the main relations that describe the dynamics of the two DGP branes embedded in a 5-dimensional background spacetime and reproduce the well-know local DGP expression for the Hubble parameters on each brane. In Section 3 we find new global solutions for the scale factors of the two branes that describe their dynamics by means of an specific form of a metric coefficient in terms of the scale factor.  Besides, we obtain the cosmological restrictions for the content of matter and energy in each brane and the relation between their equation of state parameters. In Section 4 we analyze the cosmological properties of the system for the two DGP branches which includes non-relativistic matter, cosmological constant and radiation present in one brane. We find the conditions for obtaining a universe without initial singularity, an oscillating or a recollapsing universes. Additionally, we find approximate expressions for the Hubble parameter, the energy density and the equation of state parameters for the branes in the low and high energy limit. Finally, we give our conclusions.

\section{The model} \label{Sec-2}

The action describing the dynamics of two DGP branes embedded in a 5-dimensional background spacetime is given by  
\begin{equation}
  S=-\frac{1}{2\kappa^{2}_{(5)}} \int d^{5}x \sqrt{-g_{(5)}} R_{(5)}
  + \int d^{4}x \sqrt{-g_{(4)}} ( \mathcal{L}_0 +  \mathcal{L}_c) \, ,
\end{equation}
where $g_{(5)}$, $\kappa_{(5)}^2$, and $R_{(5)}$ are the determinant of the metric $G_{AB}$, the gravitational constant and the scalar curvature in 5-dimensions respectively   
\footnote{We use the convention  
  $c=\hbar=1$, and the indices take the values  $\mu,\nu = 0,1,2,3$, $i,j=1,2,3$, and
  $A,B = 0,1,2,3,4$.}.The branes are located in  $y=0$, y $y=y_{c}$ and their corresponding Lagrangians are denoted by 
$\mathcal{L}_0$, $\mathcal{L}_c$ which include the scalar curvature in 4-dimensions and perfect fluids matter fields. The induced metric on the branes is  $g_{\mu \nu} = G_{\mu\nu}(x,y=0,y_c)$ and its determinant is denoted by $g_{(4)}$. 

The general form of the 5-dimensional metric which is consistent with $Z_2$ symmetry in the five coordinate $y$ ($(x^{\mu},y) \rightarrow (x^{\mu},-y)$)is 
\begin{equation}
  ds^{2} = - n^{2}(t,|y|) dt^{2} + a^{2}(t,|y|) g_{i j} dx^{i} dx^{j}
  +b^{2}(t,|y|) dy^{2} \,. \label{metrica}
\end{equation}
Considering that the extra dimension coordinate $y$ is compactified and satisfied $(x^{\mu},y)\to(x^{\mu},y+2iy_{c}), \, i=1,2,\ldots$ the coefficients in the metric $a(t,\left|y\right|)$,
$n(t,\left|y\right|)$ and $b(t,\left|y\right|)$ satisfy the following conditions
\begin{subequations}
\label{eq:condicion}
  \begin{eqnarray}
    \left[F^{\prime} \right]_{0} = 2F^\prime |_{y=0+} \, , \quad
    \left[F^{\prime}\right]_{c} = -2F^\prime |_{y=y_{c}-} \,
    ,  \label{condicion1} \\	
    F^{\prime \prime} = \frac{d^{2} F(t,\left| y \right|)}{d\left| y
      \right|^{2}} + \left[ F^\prime\right]_{0} \delta(y) + \left[
      F^\prime \right]_{c} \delta(y-y_{c}) \, , \label{condicion3}
\end{eqnarray}
\end{subequations}
where the prime represents the derivative with respect to $y$ coordinate, the brackets stand for the discontinuity of the first derivative at the positions $y=0$ y $y=y_{c}$ and $F$ is a general function that satisfies the last conditions. The subindex $c$ will be employed for quantities evaluated in $y =
y_{c}$ and the subindex  $0$ will be used for quantities evaluated in  $y = 0$. The Eq.~(\ref{condicion3}) was calculated under the following suppositions  $d \left|y \right| /dy = 1$ and  $d^{2} \left| y
\right|/dy^{2} = 2\delta(y) -2 \delta(y-y_{c})$, for $y \in [ 0,y_c
]$. The former relations are obtained because the introduction of the $Z_2$ symmetry on each brane. The existence of delta distributions in the second derivatives is adequate due to the fact that the Einstein equations involve second derivatives of the metric when the Einstein tensor is calculated  and the energy momentum tensor for the branes also includes delta distributions \cite{Wang:2008zzr}. 

The 5-dimensional energy momentum tensor for the system can be written as:
\begin{eqnarray}
  \label{tme} \tilde{T}^{A}_B&=&
  -\frac{\Lambda_{5}}{\kappa^{2}_{(5)}} g^{A}_B  +
  \frac{\delta(y)}{b_{0}} \mathrm{diag} (-\rho_{0},p_{0},p_{0},p_{0},0) - \frac{\delta(y)}{b_{0}\kappa^{'}} \left( R^{\mu } _{\nu}-\frac{ 1 }{ 2 } { g }^{\mu} _{\nu}R \right)e_\mu ^A e^\nu _B
  \nonumber\\
  && + \frac{\delta(y-y_{c})}{b_{c}}
  \mathrm{diag} (-\rho_{c},p_{c},p_{c},p_{c},0)-\frac{\delta(y-y_{c})}{ b_{c} \kappa^{'} } \left( R^{\mu } _{\nu}-\frac{ 1 }{ 2 } { g }^{\mu} _{\nu}R \right)e_\mu ^A e^\nu _B,
  \label{energy5d}
\end{eqnarray}
where the first term contains the 5-dimensional cosmological constant $\Lambda_5$, the second and third terms describe the perfect fluid and the contribution to the energy momentum tensor of the 4-dimensional intrinsic curvature of the brane located in $y=0$ where $R^{\mu\nu}$ and $R$ are the Ricci tensor and the scalar curvature constructed with the induced metric $g_{\mu\nu}$ on the brane. The tangent vectors to the branes are $e_\mu ^A= \frac{\partial x^A}{\partial x^\mu}= \delta^A_\mu$ where $X^A=X^A(x^\mu)$ are the embedding functions of the branes situated in $y=0$ and $y=y_c$.  The fourth and fifth terms describe the corresponding quantities for the brane located in $y=y_c$ and $\kappa'$ is the gravitational constant in 4-dimensions.
Using the conservation of the energy momentum tensor, $\nabla_{A}
\tilde{T}^{A}_{B} =0$, the energy and the pressure on each brane should satisfied the following well-known equations:
\begin{subequations}
  \begin{eqnarray}
    \label{conservacion}
    \dot{\rho}_{0} + 3(p_{0} + \rho_{0}) \frac{\dot{a}_{0}}{a_{0}} =0 \,
    , \\
    \dot{\rho}_{c} + 3(p_{c} + \rho_{c}) \frac{\dot{a}_{c}}{a_{c}} =0 \,
    .\end{eqnarray}
\end{subequations}
The dot represents the derivative with respect to time and we suppose a relation between the energy density $\rho$ and the pressure $p$ for each brane located in $y= y_0$ and $y=y_c$ through the following expressions $p_{0}= w_{0}\rho_{0}$, and $p_{c}= w_{c}\rho_{c}$.
In the next sections we will consider specific cases for the equation of state parameter $w_0$ and analyze the behavior of $w_c$.

The non vanishing components of the 5-dimensional Einstein tensor are \\
\begin{subequations}
    \label{einstein}
  \begin{eqnarray}
    \label{einstein0}
    \tilde{G}_{00} &=& 3\frac{\dot{a}}{a} \left( \frac{\dot{a}}{a} +
      \frac{\dot{b}}{b} \right) -3 \frac{n^{2}}{b^{2}}
    \left[\frac{a''}{a} + \frac{a'}{a} \left( \frac{a'}{a} -
        \frac{b'}{b} \right) \right] 
    + 3k \frac{n^2}{a^2} \, , \\
    \label{einstein1}
  \tilde{G}_{ij} &=& \frac{a^{2}}{b^{2}} \delta_{ij}
  \left\{\frac{a'}{a} \left( \frac{a'}{a} + 2\frac{n'}{n} \right) -
    \frac{b'}{b} \left( \frac{n'}{n} +2 \frac{a'}{a} \right) \right\}
  \nonumber \\
  &+& \frac{a^{2}}{b^{2}} \delta_{ij} \left\{ 2\frac{a''}{a} +
    \frac{n''}{n} \right\} + \frac{a^{2}}{n^{2}} \delta_{ij} \left\{
    \frac{\dot{a}}{a} \left( -\frac{\dot{a}}{a} + 2\frac{\dot{n}}{n}
    \right) \right\} \nonumber \\
  &+& \frac{a^{2}}{n^{2}} \delta_{ij} \left\{ -2\frac{\ddot{a}}{a} +
    \frac{\dot{b}}{b} \left( -2 \frac{\dot{a}}{a} + \frac{\dot{n}}{n}
    \right) - \frac{\ddot{b}}{b} \right\} -k \delta_{ij} \, , \\
  \label{einstein2}
  \tilde{G}_{04} &=& 3\left( \frac{\dot{a}}{a} \frac{n'}{n} +
    \frac{\dot{b}}{b} \frac{a'}{a} - \frac{\dot{a}'}{a} \right) \, , \\
  \label{einstein3}
  \tilde{G}_{44} &=& 3\frac{a'}{a} \left( \frac{a'}{a} + \frac{n'}{n}
  \right) -3 \frac{b^{2}}{n^{2}} \left[ \frac{\ddot{a}}{a} +
    \frac{\dot{a}}{a} \left( \frac{\dot{a}}{a} - \frac{\dot{n}}{n}
    \right) \right]
  - 3k \frac{b^2}{a^2} \, ,
\end{eqnarray}
\end{subequations}
which are fundamental to obtain the global behavior of the two DGP brane system embedded in a 5-dimensional background spacetime with a compactified extra dimension.  

By means of the Israel junction conditions \cite{PhysRevD.43.1129} we can express the energy density in terms of the discontinuity of the derivative of the metric in the extra coordinate $y$. The junction conditions for the brane in $y=0$ allow to write the discontinuity of the metric coefficients in terms of the energy and the pressure:
\begin{subequations}
\label{eq:saltov1}
  \begin{eqnarray}
    \label{salto1}
    \frac{\left[ a^\prime \right]_{0}}{a_{0} b_{0}} &=&
    -\frac{\kappa^{2}_{(5)}}{3} \rho_{0}+2r_0 \left(\mathcal{H}_0^{ 2 } + \frac{k}{a_0 ^2} \right) \, , \\
    \label{salto2}
    \frac{\left[ n^\prime \right]_{0}}{n_{0} b_{0}} &=&
    \frac{\kappa^{2}_{(5)}}{3} (3p_{0} + 2 \rho_{0} )+2r_0\left( \mathcal{H}_0^2+2\frac{\dot{\mathcal{H}_0}}{n_0} - \frac{k}{a_0 ^2}\right) \, .
  \end{eqnarray}
\end{subequations}
and for the brane in $y=y_{c}$ the analogous condition is, 
\begin{subequations}
\label{eq:saltov2}
  \begin{eqnarray}
    \label{salto3}
    \frac{\left[ a^\prime \right]_{c}}{a_{c} b_{c}} &=&
    -\frac{\kappa^{2}_{(5)}}{3} \rho_{c}+2r_0 \left(\mathcal{H}_c^{ 2 } + \frac{k}{a_c ^2} \right)\, , \\
    \label{salto4}
    \frac{\left[ n^\prime \right]_{c}}{n_{c} b_{c}} &=&
    \frac{\kappa^{2}_{(5)}}{3} (3p_{c} +2\rho_{c} )+2r_0\left( \mathcal{H}_c^2+2\frac{\dot{\mathcal{H}_c}}{n_c} - \frac{k}{a_c ^2} \right) \, ,
  \end{eqnarray}
\end{subequations}
where $\mathcal{H}=\frac{\dot{a}}{na}$ is a term related with the Hubble parameter which it can be interpreted as the Hubble parameter for an observer on the brane once it is evaluated in $y=0, y_c$, $r_0=\kappa_{(5)}^2/2\kappa'=M_{(4)}^2/2M_{(5)}^3$ is the crossover scale between the 4 and 5-dimensional gravity regimes and $M_{(4)}$, $M_{(5)}$ are the Planck mass in 4 and 5-dimensions respectively. From the expression of the energy momentum tensor (\ref{energy5d}) it is convenient to define the energy density $\rho_{curv}$ and pressure $p_{curv}$ associated with the intrinsic curvature term of the brane as \cite{Deffayet:2000uy}
\begin{equation}
 \rho_{curv} = -\frac{6r_0}{\kappa^2 _{(5)}}\left( \mathcal{H}^2 + \frac{k}{a^2} \right),
\end{equation}
\begin{equation}
 p_{curv} = \frac{2r_0}{\kappa^2 _{(5)}}\left( 3\mathcal{H}^2 + 2\frac{\dot{\mathcal{H}}}{n} + \frac{k}{a^2} \right) ,
\end{equation}
where these relations are evaluated in the branes located at $y=y_{0}$  and $y=y_{c}$. This ``perfect fluid'' satisfies the fluid equation ${\dot{\rho}}_{curv} + 3H(\rho_{curv} + p_{curv})=0$. The energy density associated with the brane intrinsic curvature term is always negative for $k=0$ and $k=1$. 

\subsection{Exact solutions}

In the following we will obtain exact solutions for the equations of motion by means of the following procedure. Considering Eq.~(\ref{einstein2}):
$\tilde{G}_{04} = \kappa^{2}_{(5)} \tilde{T}_{04} = 0$ (this relation means that the matter flux along the fifth dimension is null) we have
\begin{equation}
  \label{0-5}
  \frac{\dot{b}}{b} = \frac{n}{a^\prime} \left[ \frac{\dot{a}}{n}
  \right]^\prime \, .
\end{equation}
The former equation has a straightforward solution for the static case,  i.e. $\dot{b} =
0$ and it can be chosen $b =1$. In fact this kind of solution was obtain in \cite{Binetruy:1999hy,Mukohyama:1999qx}. If we look for a non-static solution, it is possible to write the metric coefficient $b$ which describes the fifth dimension in terms of a general function of the scale factor. In fact, we can integrate
Eq.~(\ref{0-5}) which gives the general relation,
\begin{equation}
  \label{eq:ansatz}
  b = f(a) \quad \Rightarrow \quad \dot{a} = n \, b \, \alpha(t) \, ,
\end{equation}
where  $f(a)$ is a general function of the scale factor $a$, and
$\alpha(t)$ is an arbitrary function of time. A general form for $f(a)$ describes a non-static internal dimension \cite{Feranie:2010th}. We can recover the static case if we set $b=1$.

If we now integrate Eq.~(\ref{einstein0}) in the $y$ coordinate, we have
\begin{equation}
  \label{friedmann3}
  \left( \frac{a^\prime}{ab} \right)^2 - \left( \frac{\dot{a}}{an}
  \right)^2 = k a^{-2} - \frac{\Lambda_{5}}{6} + C_{_{DR}} a^{-4} \, ,
\end{equation}
where $C_{_{DR}}$ is a constant which gives rise a contribution of a kind of dark radiation \cite{Maartens:2010ar}. 

For the sake of simplicity we consider the case when $k$, the dark radiation  $C_{DR}$, and the 5-dimensional cosmological constant are zero, then we have the following solution  $a^\prime = b^2 \alpha(t)$.

We propose the next form of the function $f$ in terms of the power law of the scale factor $f(a)=a^m$, and the general form of the scale factor is 
\begin{equation}
  \label{eq:1}
  a(t,y) = a_0 \left[
    1 + (1-2m) y \alpha \, a^{2m-1}_0 \right]^{1/(1-2m)} \, ,
\end{equation}
where $a_0$ is the scale factor that depends on time in $y=0$. The metric function $n$ is obtained in terms of the scale factor and its derivatives. The solution for $m=0$ was calculated in \cite{Deffayet:2000uy}.

\subsection{DGP solutions}

By means of the junctions conditions (\ref{salto1}), (\ref{salto3}), the relations (\ref{condicion1}) and Eq. (\ref{friedmann3}) we find the following expressions:
\begin{equation}
 \epsilon|\mathcal{H}_0| = r_0 \mathcal{H}_0 ^2 - \frac{\kappa^{2}_{(5)} }{6}\rho_0 ,\ \ \ \ \ \epsilon'|\mathcal{H}_c| = r_0 \mathcal{H}_c ^2 - \frac{\kappa^{2}_{(5)} }{6} \rho_c ,
\end{equation}
where $\epsilon=\pm 1$ is the sign of $[a']_0$ and $\epsilon'=\pm 1$ is the sign of $[a']_c$. The former expressions can be easily solved to obtain explicit relations for the Hubble parameters for each brane which are given by
\begin{equation}
\label{expansion rate}
\mid\mathcal{H}_{0}\mid=\frac{\epsilon+\sqrt{1+\frac{\rho_{0}}{\rho_{D}}}}{2r_{0}},\ \ \ \ \     \mid\mathcal{H}_{c}\mid=\frac{\epsilon'\pm\sqrt{1+\frac{\rho_{c}}{\rho_{D}}}}{2r_{0}} ,
\end{equation}
where $\rho_{D}^{-1}=\frac{2r_{0}k_{(5)}^{2}}{3}$ and it is assume $\rho_0 \geq 0$. It is important to notice that $n_0$ and $n_c$ are calculated from Eq.~(\ref{eq:ansatz}) therefore we only need to know the explicit form of $a_0$ to obtain a complete solution of the metric functions. The well known self-accelerating branch for the visible brane corresponds to the value $\epsilon=1$ and the normal branch to $\epsilon=-1$. For the brane located at $y=y_c$ there is an extra sign possibility in front of the squared root due to the fact that the energy density $\rho_c$ could be negative and it is related to $\rho_0$ by means of topological restrictions as we will see later. In Section IV we are going to analyze the behavior of the cosmological parameters of the two branes which are related through topological restrictions. 

\section{Global solutions}

Using the mathematical properties of the brane metric functions given by Eqs. (\ref{condicion1}) and (\ref{condicion3}) it is possible to find general solutions by taking into account the boundary conditions. For example, the junction condition ~(\ref{salto1}) can be expressed in the following way
\begin{equation}
  \frac{\left[ a^\prime \right]_{0}}{a_{0} b_{0}} = \frac{2
    b^2_0\alpha}{a_0 b_0} = 2 \alpha a^{2m-1}_0 b^{-1}_0 \, ,
\end{equation}
Considering similar equations for the metric functions, we find the complete solutions which include the matter content on the brane situated at $y=0$:
\begin{subequations}
\label{eq:solucion}
  \begin{eqnarray}
    \label{solucion1}
    a(t,y) &=& a_{0} \left[ 1 + (2m-1)\left(\frac{ \kappa^2_{(5)}} {6} {\rho}_0- r_{0}\mathcal{H}_{0}^{2}\right)  a^m_0 y \right]^{\frac{1}{(1-2m)}} \\	
    \label{solucion2}	
    n(t,y) &=& n_0{\left(\frac{a}{a_0}\right)}^{ { m } }\left[ 1 + \left( m + 2 + 3\omega_{0}
      \right) \frac{\kappa^{2}_{(5)}}{6} \rho_{0} a^m_0 y \right.
     \nonumber \\
    && \left.+r_0 a_0^m y\left((1-m)\mathcal{H}_0^2+2\frac{\dot{\mathcal{H}_0}}
{n_0}\right)\right] , \\
    \label{solucion3}
    b(t,y) &=& a^m_{0} \left[ 1 + (2m-1) \left(\frac{ \kappa^2_{(5)}} {6} {\rho}_0- r_{0}\mathcal{H}_{0}^{2}\right) a^m_0 y \right]^{\frac{m}{(1-2m)}} \ ,
\end{eqnarray}
\end{subequations}
for $m \neq 1/2$.

The former equations ~(\ref{eq:solucion}) are generalizations of the results reported in \cite{Perez:2012vx,Binetruy:1999ut,Deffayet:2000uy}. Now we have an additional parameter $r_0$ which is the ratio between the Planck mass in 5 and 4 dimensions. The solution without the additional term has been reported in ~\cite{Perez:2012vx}:
\begin{subequations}
  \begin{eqnarray}
    a(t,y) &=& a_{0} \left[ 1 + (2m-1)\left(\frac{ \kappa^2_{(5)}} {6} {\rho}_0\right)  a^m_0 y \right]^{\frac{1}{(1-2m)}}, \\	
    n(t,y) &=& n_0{\left(\frac{a}{a_0}\right)}^{ { m } }\left[ 1 + \left( m + 2 + 3\omega_{0}
      \right) \frac{\kappa^{2}_{(5)}}{6} \rho_{0} a^m_0 y \right]\, , \\
    b(t,y) &=& a^m_{0} \left[ 1 + (2m-1) \left(\frac{ \kappa^2_{(5)}} {6} {\rho}_0\right) a^m_0 y \right]^{\frac{m}{(1-2m)}} \,.
\end{eqnarray}
\end{subequations}
An interesting case arise if we take  $m=1$ and $\omega_{0}=-1$ which are conformal solutions of the type founded by Randall and Sundrum:
\begin{eqnarray}
  \label{eq:rs}
      a(t,y) &=& a_{0} \left[ 1+\left(\frac{\kappa^{2}_{(5)}}{6} \rho_{0}-r_0\mathcal{H}_0^2
      \right)a_{0}y \right]^{-1}
      \nonumber \\
      &&  = \frac{a_0 n}{n_0} = b\, .
\end{eqnarray}
Although the form of the metric is analogous to the Randall-Sundrum case, the function $a_0$ can be time-dependent. To obtain the Randall-Sundrum case we need to set $\Lambda_5 \neq 0 $, $k=0$ and $r_0 =0$ since in this case we get from Eq.(\ref{conexion0}) that $\rho_0 = -\rho_c$. Finally, in the case when $m=0$ and it is chosen $n_0 =1$,  we recover the lineal solution reported in \cite{Deffayet:2000uy}. 

The boundary conditions at $y = y_0$ and $y = y_c$ must be imposed for obtaining the global solutions given in Eqs.~(\ref{eq:solucion}). For example, these expressions imply the following relation
\begin{equation}
 \frac{a_c^\prime}{b_c}= \frac{a_0^\prime}{b_0}\left(\frac{a_c}{a_0}\right)^m ,
\end{equation} 
and an analogous relation between $n_c ^\prime$ and $n_c$. Now, the derivative discontinuities given in the Eqs. ~(\ref{eq:saltov1}) and ~(\ref{eq:saltov2}), and Eq. (\ref{eq:condicion}) give rise the following relation between the total energy densities $\rho_{T0}=\rho_0 +\rho_{0curv}$ and $\rho_{Tc}=\rho_c +\rho_{ccurv}$ on each brane 
\begin{equation}
\label{reltotalenergy}
 \frac{\rho_{Tc}}{4\rho_D}=-\epsilon^\prime r_0|\mathcal{H}_c |=-\frac{\rho_{T0}}{4\rho_D}\left(\frac{a_c}{a_0}\right)^{m-1}= \epsilon r_0 |\mathcal{H}_0| \left(\frac{a_c}{a_0}\right)^{m-1}.
\end{equation}
The last equations give important physical consequences. These relations impose topological restrictions on the kind of the energy content in the hidden brane if there is a well-know physical motivated perfect fluid on the visible brane or vice versa. We can determine that $-\epsilon$, $-\epsilon'$ gives the signs of the effective total energy density $\rho_T$ on the branes as seen from the 5D bulk. Then, the self-accelerating branch is associated with a negative effective total energy as seen from bulk and the normal branch is related with a total positive energy \cite{Deffayet:2000uy}.  The relation $\epsilon = -\epsilon^\prime$ implies that the self-accelerating branch (normal branch) in the visible brane imposes the normal branch (self-accelerating branch) in  the hidden brane and vice versa. Finally, the former relations allows to write the following relation between the energy densities on each brane.
\begin{equation}
\label{conexion0}
\rho_c= 4\rho_D x \left[\epsilon+ x \right], 
\end{equation}
where we have defined
\begin{equation}
    x:= r_0\mid\mathcal{H}_0\mid \left(\frac{a_c}{a_0}\right)^{m-1} \, .
\end{equation}
From the relation given in (\ref{conexion0}) it is possible to conclude that the energy density $\rho_c$ is always positive for the self-accelerating branch $\epsilon =1$. For the normal branch, when $\epsilon = -1$, $\rho_c$ can take negative and positive values.  Using Eq. (\ref{conexion0}) and by means of the relation $(\frac{a_c}{a_0})^{m-1}|\mathcal{H}_0| = |\mathcal{H}_c|$ the positive values for $\rho_c$ corresponds to $|\mathcal{H}_c|>1/r_0$ or $r_{\mathcal{H}_c}=\frac{1}{\mathcal{H}_c}<r_0$ and negative values for $\rho_c$ corresponds when $|\mathcal{H}_c|<1/r_0$ or $r_{\mathcal{H}_c}>r_0$. The calculation of the minimum value for $\rho_c$ is straightforward and gives $\rho_c = -\rho_D$ when $x=\frac{1}{2}$ or equivalently when $|\mathcal{H}_c|= \frac{1}{2r_0}$. This result always allows real values for the Hubble parameter $\mathcal{H}_c$ given in Eq.~(\ref{expansion rate}). In addition from Eq. (\ref{conexion0}) we obtain the following relation $\sqrt{1+\frac{\rho_c}{\rho_D}}= |2r_0 |\mathcal{H}_c| +\epsilon|$ then the positive sign in front of the square root in Eq. (\ref{expansion rate}) corresponds to the case when $r^{-1}_{\mathcal{H}_c} \geq \frac{1}{2r_0}$ and the negative when $r^{-1}_{\mathcal{H}_c} \leq \frac{1}{2r_0}$.

The relation between the equation of state parameters $w_0$ and $w_c$ for each brane can be find with the help of the Eqs. (\ref{eq:saltov1}), (\ref{eq:saltov2}), and Eq. (\ref{eq:condicion}). 
After lengthy calculations, we find the following relation between the equations of state parameters on each brane:
{\small
\begin{equation}
\label{conexion2} 
w_{c} = \frac{\left(2\epsilon x(w_0+1) + w_0-1\right)(\rho_0/4r_0|\mathcal{H}_0|\rho_D)
-\epsilon\left(1 + x\sqrt{1+\rho_0/\rho_D}\right) - \left( \epsilon +x - \frac{m}{3}\left(\epsilon+2x\right)\right)\beta a_0 ^m y_c}{\left(1+\epsilon x\right)\left(\sqrt{1+\rho_0/\rho_D} + \beta \epsilon a_0 ^m y_c\right)},
\end{equation}}
where the $\beta$ function is defined as follows
\begin{equation}
     \beta := |\mathcal{H}_0|(1-m)\sqrt{1+\rho_0/\rho_D} - 3(1+w_0)\rho_0/4r_0 \rho_D \, .
\end{equation}
Finally, the relation  between the Hubble parameters can be calculated directly from the expression (\ref{solucion1}) for the scale factor and the relation
\[
\frac{d(a_c/a_0)}{dt} =\left( H_c-H_0\right)\left(\frac{a_c}{a_0}\right)\, .
\]
The last expression can be written in the following form 
\begin{equation}
\label{Hubbleconexion}
 H_c = H_0 \frac{n_c}{n_0}\left(\frac{a_c}{a_0}\right)^{m-1} ,
\end{equation}
where it gives as a consequence the relation $\mathcal{H}_c = \mathcal{H}_0 \left({a_c}/{a_0}\right)^{m-1}$ which is consistent with the expression obtained from Eq.(\ref{reltotalenergy}). In the limit of $r_0\rightarrow 0$ the Eqs. (\ref{conexion0}), (\ref{conexion2}) and (\ref{Hubbleconexion}) reproduce the corresponding equations reported in \cite{Perez:2012vx}. By means of the fluid equation for the visible brane, it is possible to prove the fluid equation in the hidden brane $\rho_c + 3H_c(1+w_c)\rho_c = 0$.
The former expressions will be very useful in order to analyze the physical consequences of the model in the next Section. 

\section{Cosmological consequences of the brane system}\label{Sec-3}
As a starting point we are going to consider the cosmology for low and high energy limits for each brane with respect to energy scale given by $\rho_D$.  We will find the energy density, the Hubble parameter and the equation of state parameter for the hidden brane in terms of the energy density of the visible for each of the two branches in the corresponding limits. In the following analysis we take our gauge choice as $n_0=1$.

\subsection{4-dimensional gravity regime}
In the high energy limit when $\frac{\rho_0}{\rho_D}\gg1$ we can write Eqs~(\ref{expansion rate}) in a first order approximation as $H_0 ^2 = \frac{\rho_0}{4r_0 ^2 \rho_D}= \frac{8\pi G_{(4)}}{3}\rho_0$ which corresponds to recover the standard 4D Friedmann equation. This limit is equivalent to the the condition $H_0 ^{-1} \ll r_0$ where the scale $r_0$  is the crossover between the 4D gravity and 5D gravity as was found by Dvali et. al. \cite{DGP} and Deffayet \cite{Deffayet:2000uy}. 
In this regime the expression of the Hubble parameter is written in the following way in the next order of approximation 
\begin{equation}
\label{eqn:33}
H_{0}^{2}\approx \frac{\rho_0}{4r_0 ^2 \rho_D} + \frac{\epsilon}{r_0}\sqrt{\frac{\rho_0}{4r_0 ^2 \rho_D}}.
\end{equation}
For the self-accelerating branch $\epsilon =1$, we observe a positive contribution to energy density which is originated due to the existence of a energy density associated with the brane intrinsic curvature. The opposite situation happens for the normal branch when  $\epsilon =-1$. 

The expression for the Hubble parameter of the hidden brane is related to the energy density component of the visible brane as follows
\begin{equation}
\label{eqn:2.13}
H_{c}^2 \approx \frac{8\pi G_{(4)}}{3}\rho_0\frac{\left[1+ \epsilon a_0^m \frac{y_c}{2r_0} \sqrt{\frac{\rho_0}{\rho_D}}\left[(1-m) - \frac{3}{2}(1+w_0)\right] \right]^2}{\left[1- \epsilon (2m-1)a_0^m \frac{y_c}{2r_0} \sqrt{\frac{\rho_0}{\rho_D}} \right]^2}.
\end{equation}
If we consider the situation when the scale factor $a_0$ is not very small and the case when $y_c$ is of the order of the Planck length then $\frac{y_c}{r_0} \approx 10^{-61}$ \cite{Xu:2013ega} and the contributions in both brackets are very close to one. This result is reinforce even when $a_0$ is too small for $m-\frac{3}{2}(1+\omega_0)>0$. In this case we have that the Hubble parameter approaches to the asymptotic value
\begin{equation}
H_{c}^2 \approx H_{0}^2 ,
\end{equation}
and the evolution of the two branes is very similar. For $m- \frac{3}{2}(1+w_0)<0$ and in the limit when the scale factor $a_0$ is very close to zero the two Hubble parameters are related by
\begin{equation}
H_{c}^2 \approx \frac{\left[(m-1) + \frac{3}{2}(1 + \omega_0)\right]^2}{\left[2m-1\right]^2} H_{0}^2.
\end{equation}
From the former expression we can see that at the beginning of the visible universe the expansion of the hidden brane is the same for both branches.
The energy density of the hidden brane is related to the one in the visible brane in the following form
\begin{equation}
\label{eqn:2.14}
\frac{\rho_{c}}{\rho_D}= 4B\left(\epsilon + B\right),
\end{equation}
where we have defined $B\approx \sqrt{\frac{\rho_{0}}{4\rho_D}}  \left[1-\epsilon(2m-1)a_0^m\frac{y_c}{2r_0}\sqrt{\frac{\rho_0}{\rho_D}} \right]^{\frac{m-1}{1-2m}} $. 

The equation of state parameter has the following form in the high energy regime:
\begin{equation}
\omega_c = \frac{\frac{1}{2}(w_0-1) + \epsilon w_0 B +\left(\epsilon + B - \frac{m}{3}\left(\epsilon +2B\right)\right)(1+2m+3w_0) \frac{y_c}{4r_0}a_0^m\sqrt{\frac{\rho_0}{\rho_D}}}{(1+\epsilon B)\left( 1 -\epsilon(1+2m+3w_0)\frac{y_c}{4r_0}a_0^m\sqrt{\frac{\rho_0}{\rho_D}} \right)}.
\end{equation}
In the case when $m-\frac{3}{2}(1+\omega_0)>0$, and the scale factor is very small, it is possible to have the following asymptotic relations $\rho_{c} \simeq \rho_0$ and $w_c \simeq w_0$ together with $H_c \simeq \mathcal{H}_c \simeq \mathcal{H}_0=H_0$ which are consistent with the result obtained before for the Hubble parameters in the same case and the standard cosmology is recovered on both branes. The case $m-\frac{3}{2}(1+\omega_0)< 0$ is discussed in subsection {\bf D}.

\subsection{5-dimensional gravity regime and late time cosmology}

In the normal branch case for $\epsilon = -1$ and the low energy limit  when $|\frac{\rho_0}{\rho_D}|\ll 1$, we can write Eq. ~(\ref{expansion rate}) at first order approximation as $\mathcal{H}_0 \approx \frac{\rho_0}{4r_0\rho_D}$. From the last relation this limit is equivalent to $H_0 ^{-1} \gg r_0$ and cosmology enters to a 5-dimensional gravity regime where the Friedmann equation can be written in the following form
\begin{equation}
\label{expansion rate low energy minus zero 0}
H_{0}^{2}\approx \frac{k^4_{(5)}}{36}\rho_0^2.
\end{equation}
This limit corresponds to the usual high energy limit in other brane worlds models \cite{Maartens:2010ar} where the square of the Hubble parameter is proportional to the square of the energy density instead to be proportional to the energy density as in standard cosmology. The expression for the Hubble parameter of the hidden brane is related to the Hubble parameter for the visible brane by means of Eq.~(\ref{Hubbleconexion}) in an approximate form as follows 
\begin{equation}
 \label{expansion rate low energy minus zero c}
H_{c}^2\approx H_{0}^{2}\frac{\left[1+(m+2+3w_0)\frac{\rho_0}{4\rho_D} a_0^m \frac{y_c}{r_0} \right]^2}{\left[1+(2m-1)\frac{\rho_0}{4\rho_D}a_0^m \frac{y_c}{r_0} \right]^2}.
\end{equation}
The approximate expression for the energy density of the brane located at $y=y_c$ is given by
\begin{equation}
\label{eqn:2.11}
\frac{\rho_{c}}{\rho_D} \approx 4C(C-1) ,
\end{equation}
where $C=\frac{\rho_0}{4\rho_D}\left[1+(2m-1)\frac{\rho_0}{4\rho_D} a_0^m \frac{y_c}{r_0} \right]^\frac{m-1}{1-2m}$
and the equation of the state parameter has the following approximate form
{\small
\begin{equation}
w_c \approx \frac{w_0 -C(2w_0+1)+\left(C(1+w_0)- \frac{(1+w_0)}{2}\right)\frac{\rho_0}{2\rho_D}+ \left(C-1- \frac{m}{3}\left(2C-1\right)\right)(m+2+3w_0)\frac{\rho_0}{4\rho_D} a_0^m \frac{y_c}{r_0}}{(1-C)\left(1+(m+2+3w_0)\frac{\rho_0}{4\rho_D} a_0^m \frac{y_c}{r_0}\right)}.
\end{equation}}
In the case when $m-3(1+\omega_0)<0$ we have that $\mathcal{H}^2_c \approx H_c ^2 \approx H_0 ^2$ and the evolution of the two branes is very similar for late time cosmology. Besides, in the same case, we obtain that $\rho_c \approx -\rho_0$ and $w_c \approx w_0$ which are consistent with the evolution of the branes. If $m-3(1+\omega_0)>0$ the Hubble parameter of the hidden brane approaches to the asymptotic value $H_c ^2 \approx H_0 ^2 \frac{(m+2+3w_0)^2}{(2m-1)^2}$ and the evolution of the two branes differs at late time. The behaviors of other physical quantities are discussed in the Subsection {\bf D}.

In the self-accelerating branch when $\epsilon = 1$ and the low energy limit $|\frac{\rho_0}{\rho_D}|<<1$, we can use Eqs.~(\ref{expansion rate}) and Eqs.~(\ref{conexion0}) in order to calculate the Hubble parameter in each brane and the energy density for the hidden brane as approximate expressions. The Hubble parameter for the visible brane is
\begin{equation}
\label{expansion rate low energy plus zero 0n}
H_{0}^{2}\approx\frac{1}{r_{0}^{2}} ,
\end{equation}
which it is equivalent to have a cosmological constant but it is a by product of the brane intrinsic curvature and the existence of the extra dimension. 
The Hubble parameter for the brane located at $y_c$ is related to the Hubble parameter of the visible brane  in the following form
\begin{equation}
\label{expansion rate low aenergy plus c}
H_{c}^2\approx H_{0}^{2}\frac{\left[1 + \frac{a_0^m y_c}{r_0} (1-m)\right]^2}{\left[1-(2m-1)\frac{a_0^m y_c}{r_0}\right]^{2}}.
\end{equation}
In the case of $m<0$ it is possible to perform an additional approximation for a very large scale factor $a_0$ then we have $H_c ^2 \approx H_0 ^2 \approx 1/r_0 ^2$ and the evolution of the hidden brane presents an accelerated expansion. If $m>0$ the asymptotic value of the Hubble parameter is $H_c ^2 \approx H_0 ^2\frac{(m-1)^2}{(2m-1)^2}$ and the expansion of the hidden brane is slower (faster) than the visible brane for $m>2/3$ ($m<2/3$). The behaviors of $\rho_c$ and $w_c$ are treated for this case in subsection {\bf D}.

The approximate expression for the energy density of the brane located at $y=y_c$ is given by
\begin{equation}
\label{eqn:2.10}
\frac{\rho_{c}}{\rho_D} = 4A \left[1+ A \right] ,
\end{equation}
where we define $A\approx \left[1-(2m-1)\frac{a_0^m y_c}{r_0}\right]^\frac{m-1}{1-2m}$. 
The equation of state parameter has the following approximate form:
\begin{equation}
    w_c\approx\frac{1 + A + \left( 1+A - \frac{m}{3}\left(1+2A\right)\right)(1-m)\frac{a_0^m y}{r_0}}{\left(1+ A \right)\left((m-1)\frac{a_0^m y}{r_0} -1 \right)}.
\end{equation}
If $m<0$ and for a large scale factor $a_0$ we have that $\frac{\rho_c}{\rho_D} \approx 8$ and $w_c \simeq -1$. This result is consistent with the accelerated expansion of the hidden brane since the expression for the Hubble parameter given in Eq. (\ref{expansion rate}) produces as a result that $|\mathcal{H}_c| \approx \frac{1}{r_0}$.
\subsection{ Cosmology with static extra dimension }
We consider the special case when  $m=0$ which corresponds to a static fifth dimension when $\dot{b}=0$. In the case of  a constant equation of state parameter for the visible brane $w_0$, the solution of the conservation equation ~(\ref{conservacion}) is $\rho_0=\rho_va_0^{-3(1+w_0)}$ where $\rho_v$ is a constant. The equation ~(\ref{solucion1}) for the scale factor $a_c$, the Eqs.~(\ref{expansion rate}), (\ref{Hubbleconexion}) for the Hubble parameter $H_c$ and the Eq.~(\ref{conexion0}) for the energy density $\rho_c$ can be rewritten as follows
 \begin{equation}
\label{36}
a_c=a_0\left[1+\epsilon\mid H_0\mid y_c\right] \, ,
\end{equation}
\begin{equation}
    \label{37}
    H_c^2=\frac{ \mid H_0 \mid^2}{\left[1+\epsilon \mid H_0 \mid y_c \right]^2}
  \left[ 1 - \left( 2 + 3\omega_{0}
      \right)\epsilon \mid H_0 \mid  y_c
     +r_0 y_c\left(3(1+w_0)H_0^2+2\dot{H_0}\right)\right]^2 \, ,
\end{equation}
\begin{equation}
    \label{37b}
  \frac{\rho_c}{\rho_D} = \frac{4r_0\epsilon \mid H_0 \mid}{1+\epsilon \mid H_0 \mid y_c } \left[1+\frac{r_0\epsilon \mid H_0 \mid}{1+\epsilon \mid H_0 \mid y_c} \right] \, ,
\end{equation}
where $|{H}_{0}|=\frac{\epsilon+\sqrt{1+\frac{\rho_{0}}{\rho_{D}}}}{2r_{0}}$. The equation of state parameter (\ref{conexion2}) written explicitly for the static case is not especially illuminating.
The relation (\ref{36}) between both scale factors can give raise restrictions on the evolution of the scale factor for the visible brane in the normal branch. Considering that the scale factors of the two branes must be positive we have the condition $|\mathcal{H}_0| \leq \frac{1}{y_c}$ which implies a bound from below for the scale factor if $1+w_0> 0$
\begin{equation}
a_0 \geq a_b=\left( \frac{\rho_v y_c ^2}{4r_0\rho_D (r_0 + y_c)}\right)^{1/3(1+w_0)} \, ,
\end{equation}
and a bound from above for the scale factor when $w_0 < -1$.
\begin{equation}
a_0 \leq a_r=\left( \frac{4r_0\rho_D (r_0 + y_c)}{\rho_v y_c ^2}\right)^{1/3|1+w_0|} \, .
\end{equation} 
If we take $w_0=1/3$ and $y_c$ of the order the Planck length then the minimum scale factor for the visible brane $a_b$ is of order of the  scale factor in the Planck epoch and the maximum scale factor $a_r \approx 10^{30}$ which corresponds to the very far future. These cases are related to a Big Bounce and Big Crunch cosmology respectively.

\begin{figure}[htp!]
  \includegraphics[width=0.45\textwidth]{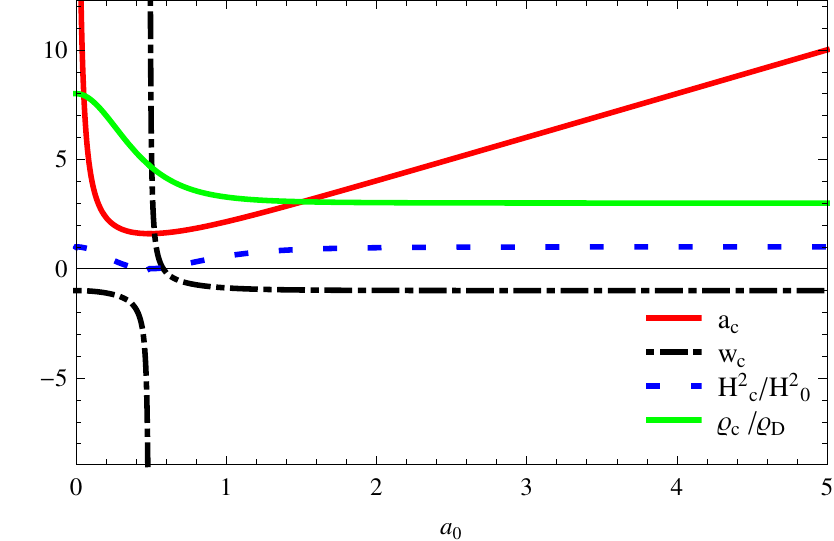}
  \includegraphics[width=0.45\textwidth]{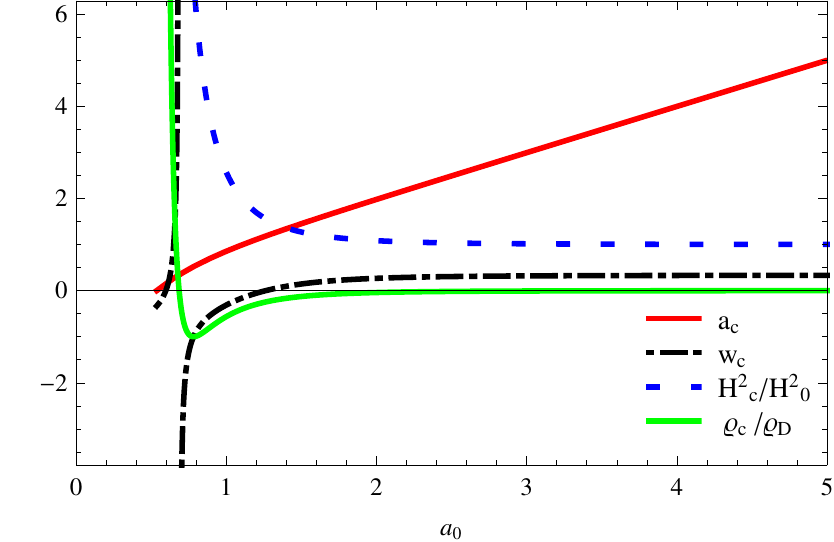}
  \caption{(left) The evolution of physical quantities as the scale factor $a_c$, the energy density $\rho_c$, the equation of state $w_c$ and the Hubble parameter $H_c$ of the hidden brane as functions of the scale factor of the visible brane $a_0$ for $w_0=1/3$ for the self-accelerating branch. (right) The same as before but for the normal branch. For simplicity and in order to appreciate the cosmological features of the system we set the following values of the parameters for all figures in the paper: $r_0=1$, $y_c=1$, $\rho_v/\rho_D=2/3$. Besides, we only plot the physical quantities in the allowed range for the scale factor $a_0$.}
  \label{figsfsw1s3}
\end{figure}

It is important to notice the divergence of the Hubble parameter and the energy density from Eqs. (\ref{37}) and (\ref{37b}) as $a_0 \rightarrow a_{b,r}$. In the self-accelerating branch there are not restrictions on $a_0$. In Fig. (\ref{figsfsw1s3}) is presented the evolution of the scale factor $a_c$, the equation of state $w_c$, the Hubble parameter $H_c$ and the energy density $\rho_c$ of the hidden brane as functions of the scale factor in the visible brane $a_0$ when we consider only radiation. For simplicity and in order to appreciate the cosmological features of the system we set the following values of the parameters for all the figures in the paper: $r_0=1$, $y_c=1$, $\rho_v/\rho_D=2/3$. Besides, we only plot the physical quantities in the allowed range for the scale factor $a_0$. As was stated earlier in the discussion of late time cosmology the evolution of the two branes is very similar since $H_c \approx H_0$ and differ in the early time cosmology. For example, in the self-accelerating branch at early times the hidden brane is contracting while the visible brane is expanding and $H_c \sim - H_0$. The asymptotic value of the equation of state parameter (\ref{conexion2}) is $w_c \approx -1$ for very small values of the scale factor since in this case $\rho_c/\rho_D \approx 4r_0/y_c(1+r_0/y_c)$ and $|\mathcal{H}_c| \approx 1/y_c$. The contraction of the hidden brane stops when it reaches its minimum value for $H_c=0$ and the equation of state parameter diverges due to the divergence of the pressure at this point and then expands until at late times the evolution is very similar. In fact, for very large value of the scale factors $a_0$ both branes experience an accelerated expansion since we have $\mathcal{H}_c \approx \frac{H_c}{1+y_c/r_0}\approx \frac{H_0}{1+y_c/r_0}$. The last result is consistent with the asymptotic form of the energy density $\rho_c \approx \frac{4\rho_D}{1 + yc/r_0}[1 + 1/(1+y_c/r_0)]$ and the equation of state parameter  $w_c \approx -1$. We have in this case a bounce cosmology for the hidden brane. 

In the case of the normal branch there is a bound from below for the scale factor $a_0$ of the visible brane in order to have always positive values for the scale factor $a_c$. Near the time when the hidden brane starts a big bang cosmology, the expansion of the hidden brane is larger than the visible brane since $H_c ^2 \gg H_0 ^2$ ($H_c$ diverges at the lower bound of $a_0$) and at late times the evolution is quite similar since $|\mathcal{H}_c| \approx H_c \approx H_0 $, $\rho_c \approx - \rho_0$ and the asymptotic value of the equation of state parameter is $w_c \approx w_0$. The divergence of the parameter of the equation of state is a consequence that the energy density $\rho_c$ is zero near the Big Bang cosmology for the hidden brane, and when  $a_0 \rightarrow a_{b}$ the energy density diverges and the equation of state parameter takes the value $w_c=-1/3$.

All the previous cosmological behaviors are very similar when a non-relativistic matter is considered for the perfect fluid in the visible brane where $w_0 =0$. One of the main differences  comes from the energy densities, since in this case it is bigger than the relativistic one for large scale factors and the asymptotic behavior of the Hubble parameters when $H_c \approx H_0$ for the branes is reached faster for the relativistic fluid than the non relativistic case.

Another interesting case is to consider the cosmological constant equation of state where $w_0 =-1$. In this situation the scale factor of the two branes are proportional and the scale factor for the hidden brane is bigger for the self-accelerating case than for the normal branch. Since $H_0$ is constant, and by means of Eq. (\ref{36}) it is possible to show that the Hubble parameters $H_0$ and $H_c$ are equal at all times as it can be corroborated directly from (\ref{37}) and (\ref{expansion rate}) where the plus (minus) sign in front of the square root is taken for the self-accelerating (normal) branch since we have $|\mathcal{H}_c|= \frac{|H_0|}{1+\epsilon|H_0|y_c}$ and $n_c = 1+ \epsilon|H_0|y_c$. The last result can be obtained by means of Eq. (\ref{37b}) and taking into account that $\rho_c >0$ for the selfaccelerated case and $\rho_c <0$ for the normal branch. Finally, in both branches we have $w_c=-1$ which is obtained from (\ref{conexion2}) and it is consistent with the former results.

\subsection{Cosmology with dynamic extra dimensions}

Now, we are interested in the case when the fifth dimension evolve with time, in other words $b=a^m$ and $\dot{b}\neq0$. The generalizations of the scale factor in the non-static case can be written in an alternative form as 
\begin{equation}
\label{41}
a_c=a_0\left[1-(2m-1)\epsilon \mid H_0 \mid a_0^m y_c\right]^{\frac{1}{1-2m}}.
\end{equation}
This relation between the scale factors of the visible and hidden brane gives possible restrictions on the physical admissible values of $a_0$. The Hubble parameter for the hidden brane as the following explicit form
\begin{equation}
\label{Hubbleexplicit}
H_c =\frac{\left[ 1+ \frac{a_0 ^m y_c}{2r_0}\left[1-m + \epsilon\frac{2(1-m) + (2(1-m)-3(1+w_0))\rho_0/\rho_D}{2\sqrt{1+ \rho_0/\rho_D}}\right]\right]}{1-(2m-1)\epsilon\mid H_0 \mid a_0^m y_c} H_0.
\end{equation}
The expressions for the energy density and the equation of state parameter of the hidden brane are given by the Eqs.~(\ref{conexion0}) and (\ref{conexion2}). There are several possibilities for the cosmological evolution of the visible brane in terms of combinations of the parameters $m$ and $w_0$ and the selection of the branch.

\subsubsection{Self-accelerating branch}
The possible cosmological restrictions for the scale factor of the visible brane can be extracted by analyzing the behavior of $\left[1-(2m-1)\epsilon \mid H_0 \mid a_0^m y_c\right]$ for small and large values of the the scale factor $a_0$ and the conditions for having an extrema. For $m<1/2$ there are not restrictions on $a_0$ due to the fact that the two terms in the expression of the bracket are positives. Then, for $m>1/2$ it is possible to have several cosmological scenarios due to the restrictions on the scale factor $a_0$.
\begin{itemize}
 \item If $2m-3(1+w_0)<0$ we have that $a_0 ^m|{H}_0| \rightarrow \infty$ when $a_0 \rightarrow 0$ and $a_0 ^m|{H}_0| \rightarrow \infty$ for $a_0 \rightarrow \infty$. From this behavior we infer from the structure of the bracket that negative values for $a_c$ could be possible for small and large values of the scale factor $a_0$ then it could be necessary to impose a bound from below and above for the scale factor. Besides, the bracket has a maximum value. These results could correspond to the cosmological scenario of an oscillating universe.
 \item If $2m-3(1+w_0)>0$ \footnote{For $2m-3(1+w_0)=0$ the following relations hold: If $a_0 \rightarrow 0$ then $a_0 ^m|{H}_0| \rightarrow \frac{1}{2r_0}\sqrt{\frac{\rho_v}{\rho_D}}$ and for $a_0 \rightarrow \infty$ then $a_0 ^m|{H}_0| \rightarrow \infty$, therefore it is possible the have the existence of a bound from above for the scale factor.} we obtain that $a_0 ^m|{H}_0| \rightarrow 0$ for $a_0\rightarrow 0$  and $a_0 ^m|{H}_0| \rightarrow \infty$ when $a_0 \rightarrow \infty$, then for large values of the scale factor $a_0$, negative values for $a_c$ are possible and the former argument gives as a result a possible bound from above for the scale factor $a_0$. The last result is related to the cosmological situation of a Big Crunch. 
\end{itemize}
\subsubsection{Normal branch}
In the case of $m>1/2$, since the two terms in the square bracket are positive, the scale factor $a_0$ can take any positive value without any restriction. Now, for $m<1/2$ the bracket could have negative values and this condition impose restrictions on the values of the scale factor $a_0$ which produce two different cosmological scenarios.
\begin{itemize}
 \item If $2m-3(1+w_0)<0$ and $w_0 >-1$ there is a bound from below for $a_0$. The same situation happens in the case when $w_0<-1$ and $m<3(1+w_0)$. For both cases we have $a_0 ^m|{H}_0| \rightarrow \infty$ when $a_0 \rightarrow 0$ and $a_0 ^m|{H}_0| \rightarrow 0$ for $a_0 \rightarrow \infty$. From this behavior we conclude that arbitrary small values for $a_0$ are not allowed since only positive value for $a_c$ are permitted and this scenario could represent a Big Bounce cosmology.
 \item If $2m-3(1+w_0)>0$ we have a bound from above for $a_0$ for two different selections of the parameters. The first selection corresponds to $1/2>m>3(1+w_0)>0$ which gives the following restriction on the equation of state parameter $-1<w_0<-5/6$. The second selection satisfy the inequalities $w_0<-1$ and $m<1/2$ including negative values of $m$. In these two cases we have the following limits $a_0 ^m|{H}_0| \rightarrow 0$ when $a_0\rightarrow 0$  and $a_0 ^m|{H}_0| \rightarrow \infty$ for $a_0 \rightarrow \infty$. This situation could be related to a Big Crunch cosmology.
\end{itemize} 
Even if the condition $m<1/2$ for having a possible restriction on the scale factor $a_0$ is satisfied there are two cases for which the minimum value of $\left[1-(2m-1)\epsilon \mid H_0 \mid a_0^m y_c\right]$ is close to one when $y_c/r_0 \ll 1$ and there is not in fact any restriction for $a_0$. The first case corresponds to $0>3(1+w_0)/2\geq m>3(1+w_0)$ and the second one to $0<3(1+w_0)/2\leq m<3(1+w_0)$. However, in the opposite case when $y_c/r_0 \gg 1$, the minimum value of the square bracket can be negative and the scenarios of Big Bounce or Big Crunch could be present.

In order to analyze the cosmological characteristics of the two-brane set-up we are going to study the system for specific equations of state in the visible brane which correspond to well motivated physical situations. We consider, for example, non-relativistic matter in the visible where $w_0=0$ since the radiation case presents an analogous behavior as was mentioned before. 

\begin{figure}[htp!]
  \includegraphics[width=0.45\textwidth]{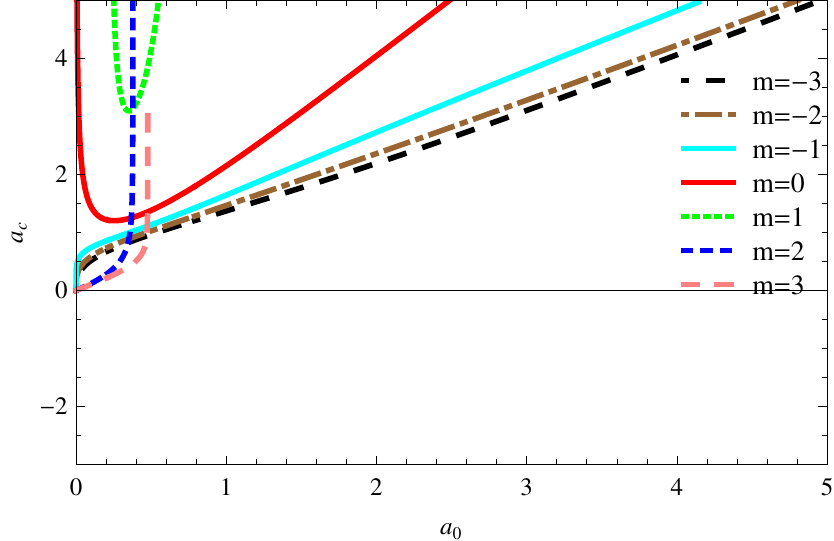}
  \includegraphics[width=0.45\textwidth]{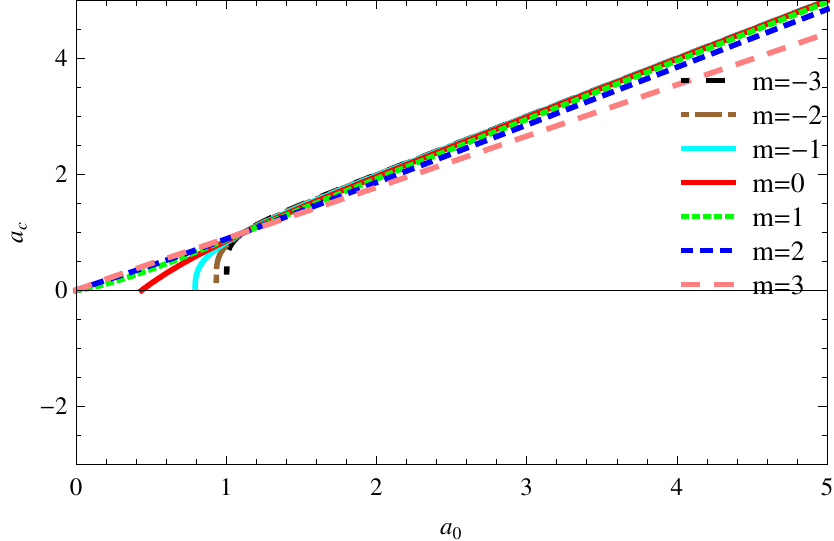}
  \caption{(left) The evolution of the scale factor $a_c$ of the hidden brane as function of the scale factor of the visible brane $a_0$ with $w_0=0$ for the self-accelerating branch. (right) The same as before but for the normal branch.}
  \label{scalefactorw0vm}
\end{figure}

\begin{figure}[htp!]
  \includegraphics[width=0.45\textwidth]{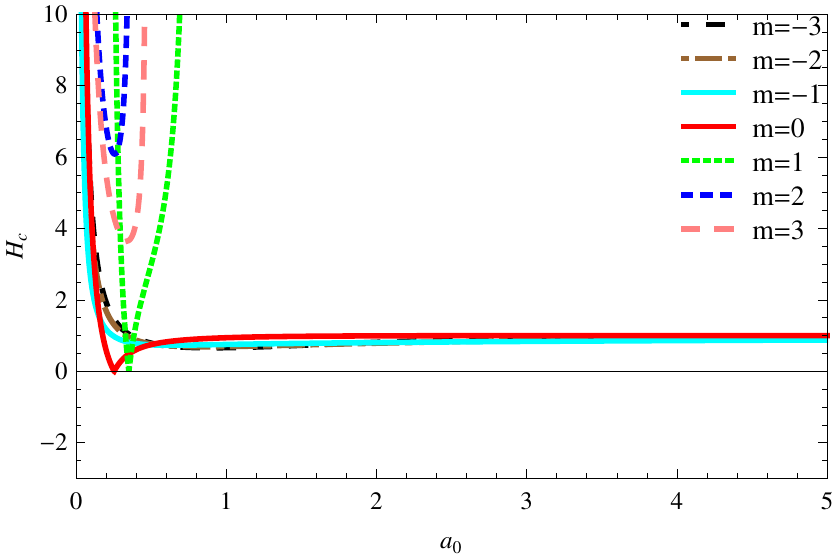}
  \includegraphics[width=0.45\textwidth]{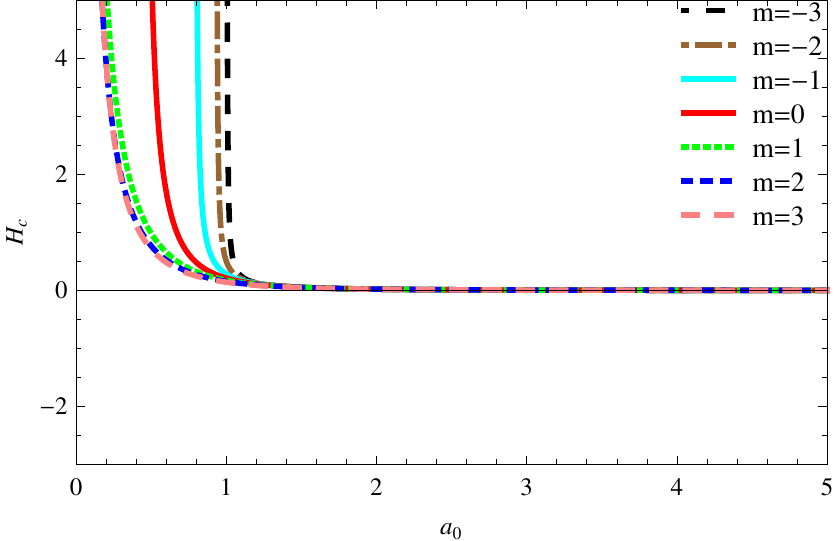}
  \caption{(left) The evolution of the Hubble parameter $H_c$ of the hidden brane as function of the scale factor of the visible brane $a_0$ with $w_0=0$ for the self-accelerating branch. (right) The same as before but for the normal branch.}
  \label{hubblew0}
\end{figure}

For the self-accelerating branch we present the behavior of the scale factor of the hidden brane as a function of the scale factor of the visible brane in Fig. (\ref{scalefactorw0vm}) for integer values of $m$. For negative values of $m$ there are not any restrictions on the possible values of $a_c$. The two branes could start with a big bang and expand together. In fact, from Fig. (\ref{hubblew0}), the values of the the Hubble parameter $H_c$ diverges when the scale factor goes to zero as it was expected according to the discussion presented in the cosmology of 4-dimensional gravity regime since $H_c\sim H_0 \rightarrow \infty$. For large values of the scale factor and, since the conditions $1+w_0>0$, $m \leq 0$ are satisfied, the evolution of the two branes is very similar and they present an accelerated expansion as it was established before in the results obtained in the study of late time cosmology. For $m=1$ the condition $1/2 <m<3(1+w_0)/2$ is satisfied and there is an oscillating universe behavior for the visible brane as it can be seen in Fig. (\ref{scalefactorw0vm}). For the oscillating universe the Hubble parameter diverges in the turning points (see Fig.  (\ref{hubblew0})) as it can be confirmed from Eq. (\ref{Hubbleexplicit}). For $m \geq 3/2$ we have a bound from above for the scale factor as it was stated in the previous results and in this case both branes can start with a Big Bang where the Hubble parameter diverges however the expansion of the hidden brane is slower than the case with $m<0$ for small values of the scale factor. In the maximum value of the scale factor the Hubble parameter also diverges. Finally, the Hubble parameter of the hidden brane is zero for the minimum value of $a_c$ as it occurs for the cases $m=0,1$ since the Hubble parameter has the alternative expression $H_c = \frac{H_0 a_0}{a_c}\frac{da_c}{da_0}$.

\begin{figure}[htp!]
  \includegraphics[width=0.45\textwidth]{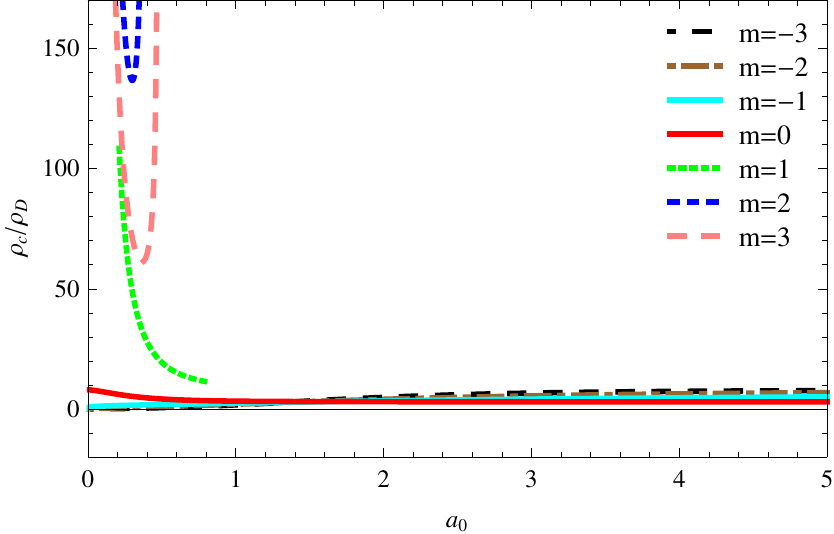}
  \includegraphics[width=0.45\textwidth]{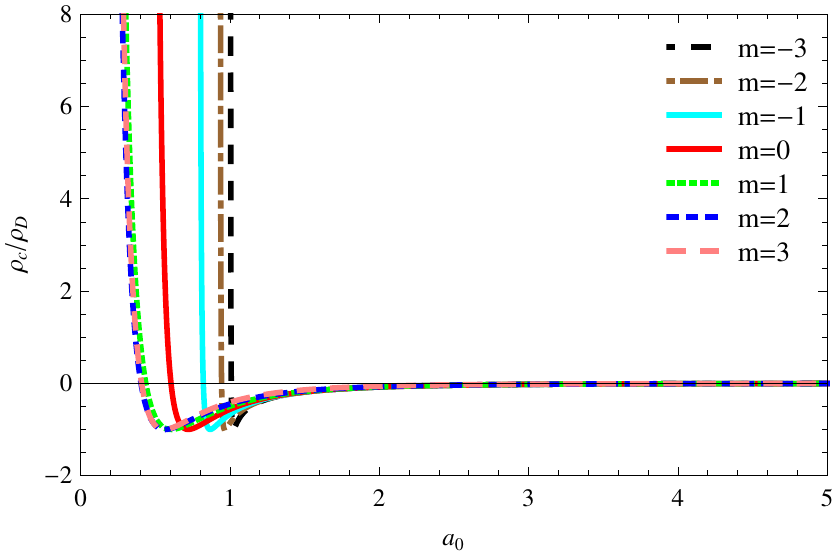}
  \caption{(left) The evolution of the energy density $\rho_c$ of the hidden brane as function of the scale factor of the visible brane $a_0$ with $w_0=0$ for the self-accelerating branch. (right) The same as before but for the normal branch.}
  \label{rhocw0}
\end{figure}

It is possible to extract the following results analyzing the behavior of the energy density $\rho_c$ from Eq.(\ref{conexion0}). The energy density $\rho_c$ tends to infinity when $m\geq 3/2$ for $a_0\rightarrow 0$ (see Fig. (\ref{rhocw0})). For $1/2<m<3/2$ we have an oscillating universe and the value $a_0 = 0$ is not allowed. In the turning points we have several possibilities for the values of the energy densities. For $1/2 < m<1$ the energy density is zero and for $1<m<3/2$ it tends to infinity. The case $m=1$ is very special since the value of the energy density is a constant given by $\rho_c = 4\rho_D r_0|\mathcal{H}_0|[1 +r_0 \mathcal{H}_0]$ evaluated at the corresponding turning points $a_0 = a_t$. For $0<m<1/2$ there is not any restriction for the scale factor and it can take the value zero where the energy density $\rho_c =0$.  In the case when $- 1/2 <m<0$ the energy density tends to infinity and for $m<-1/2$ the energy density is null. Finally for $m=-1/2$ the energy density is $\rho_c =2 \rho_D \left( \frac{r_0 ^3}{y_c ^3}\sqrt{\frac{\rho_v}{\rho_D}}\right)^{1/4} \left[ 1+ \left( 2^{-4}\frac{r_0 ^3}{y_c ^3}\sqrt{\frac{\rho_v}{\rho_D}}\right)^{1/4}\right]$ when $a=0$. For very large values of the scale factor the energy density $\rho_c$ tends to the following values. For $m<0$ we have $\rho_c \approx 8\rho_D$ and for $0<m<1/2$ the energy density is zero when the scale factor $a_0$ goes to infinity (see Fig.\ref{rhocw0}). 

\begin{figure}[htp!]
  \includegraphics[width=0.45\textwidth]{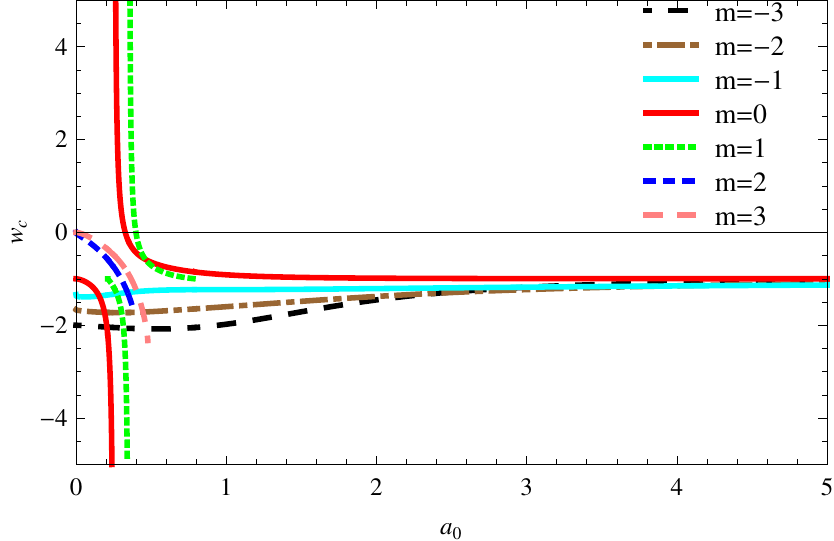}
  \includegraphics[width=0.45\textwidth]{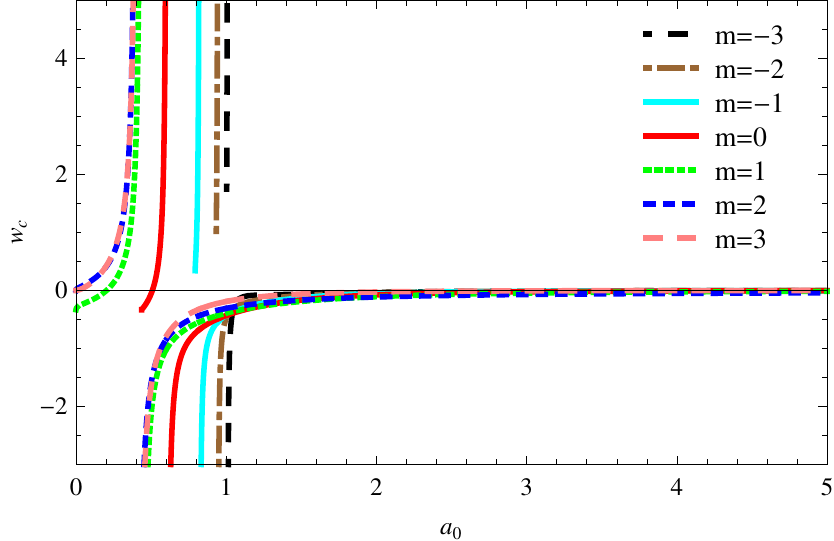}
  \caption{(left) The evolution of the equation of state parameter  $w_c$ of the hidden brane as function of the scale factor in the visible brane $a_0$ with $w_0=0$ for the self-accelerating branch. (right) The same as before but for the normal branch. We only plot $w_c$ in the allowed range for the values of the scale factor $a_0$.}
  \label{wcparameterw0}
\end{figure}

In the high energy limit the equation of state parameter has the following expressions for $a_0 \rightarrow 0$: If $0<m<1/2$, then $w_c\rightarrow -1 +m/3$, for  $-1/2<m<0$, then $w_c\rightarrow -1 +2m/3$, and $m<-1/2$ then $w_c\rightarrow -1 +m/3$. For $3/2>m>1/2$ the cosmological scenario corresponds to  oscillating universe where there are bounds from above and below for the scale factors. In those bounds the equation of state parameter takes the following values: if $1/2<m<1$ then $w_c =-\frac{m+2}{3}$, in the case $m=1$ we have $w_c=-1$ and for $m>1$ we get $w_c= -\frac{2m+1}{3}$. In the low energy limit for $0<m<1/2$ and when the scale factor $a_0$ is very large the equation of state parameter takes the asymptotic limit $w_c \approx -1 +m/3$ and in the case of $m<0$ it goes to $w_c \approx -1$. The equation of state parameter $w_c$ diverges for $m=0,1$ due to the divergence of the pressure $P_c$ (see Fig. (\ref{wcparameterw0})).

In the case of the normal branch we have a bound from below for $a_c$ for $m<1/2$ as it can be seen in Fig. (\ref{scalefactorw0vm}). For $m>1/2$ there is not any restriction for the scale factor and both branes start from a Big Bang. The expansion of hidden brane for positive $m$ is slower than for the negative values for larger values of the scale factor. The Hubble parameter diverges in the values when $a_c =0$ which corresponds to the minimum value of $a_0$ or when $a_0 =0$ for $m>1/2$. For large values of the scale factor $a_0$ the Hubble parameter of the hidden brane tends to zero as it was calculated in the late time cosmology analysis.

The energy density $\rho_c$ can take negative values and diverges at the minimum scale factor $a_0$ and for $a_c=0$ in the case when $m>1/2$. For very large values of the scale factor $a_0$ the energy density tends to zero. 

The equation of state parameter in the high energy limit, when there is not restriction on the scale factor $a_c$, takes the asymptotic value $w_c \rightarrow -1 + 2m/3$ for $3/2>m>1/2$ and $w_c \rightarrow w_0$ for  $m>3/2$ when $a\rightarrow 0$. In the low energy limit, for very large scale factor, the equation of state parameter takes the value $w_c \approx w_0=0$. The equation of state parameter diverges in $\rho_c=0$ for all the values selected in Fig. (\ref{wcparameterw0}).

\begin{figure}[htp!]
  \includegraphics[width=0.45\textwidth]{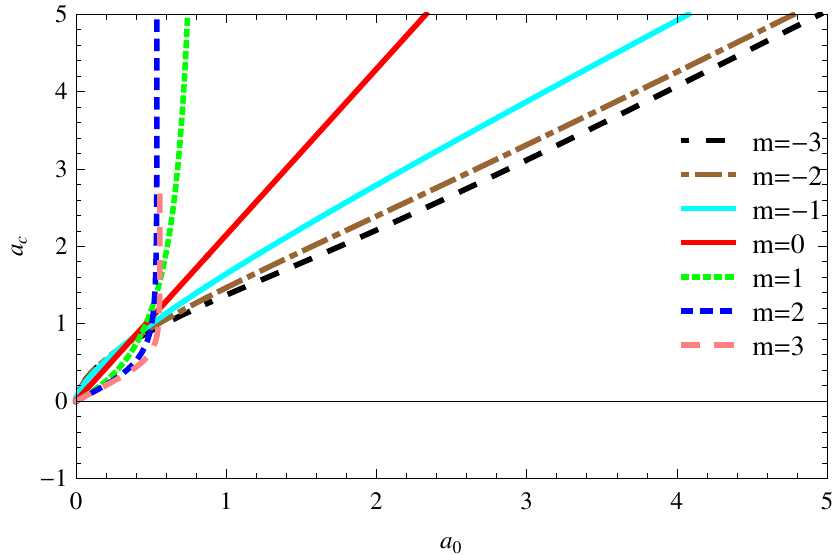}
  \includegraphics[width=0.45\textwidth]{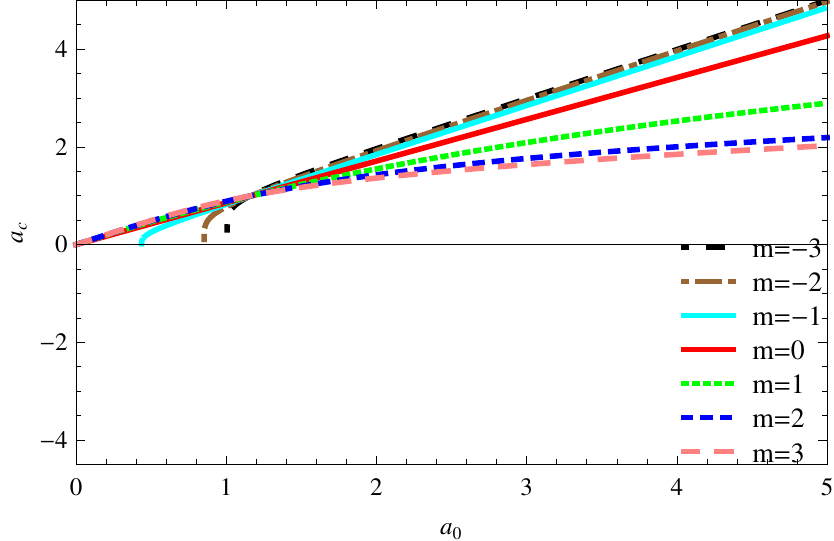}
  \caption{(left) The evolution of the scale factor $a_c$ of the hidden brane as function of the scale factor of the visible brane $a_0$ with $w_0=-1$ for the self-accelerating branch. (right) The same as before but for the normal branch}
  \label{scalefactorwm1vm}
\end{figure}

\begin{figure}[htp!]
  \includegraphics[width=0.45\textwidth]{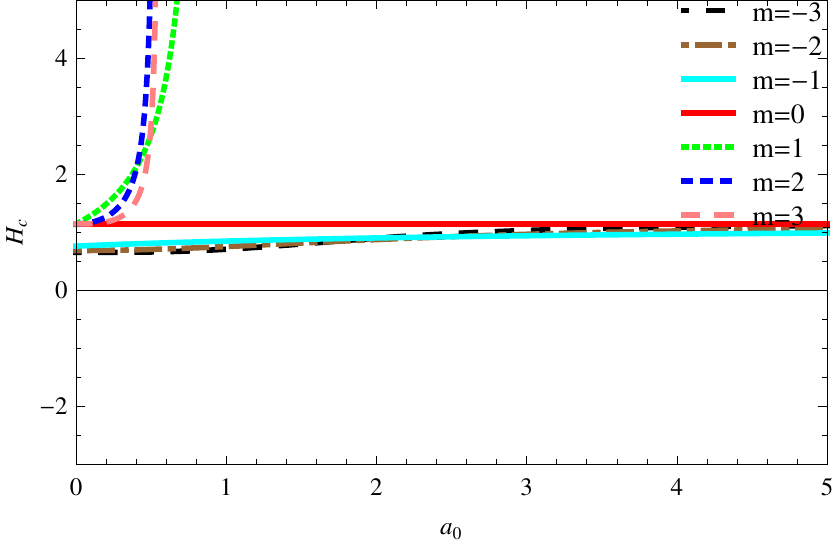}
  \includegraphics[width=0.45\textwidth]{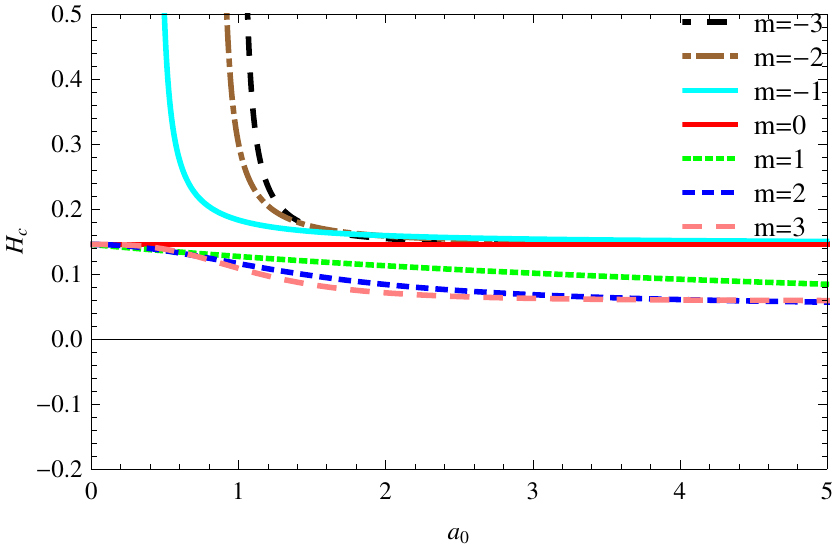}
  \caption{(left) The evolution of the Hubble parameter $H_c$ of the hidden brane as function of the normalized scale factor in the visible brane $a_0$ with $w_0=-1$ for the self-accelerating branch. (right) The same as before but for the normal branch.}
  \label{hubblewm1}
\end{figure}

In the case when we consider the equation of state for a cosmological constant $w_0=-1$ the following characteristics of the cosmological features can be extracted from the previous analysis. In the self-accelerating branch we have that only the scenario for a bound from above for the scale factor is present since the conditions for having an oscillating universe are not possible to satisfy at the same time (it is required that $m>1/2$ and $m<0$). If $m>1/2$ we have a bound from above for the scale factor $a_0$ and for $m<1/2$ we do not have restrictions at late time (see Fig. (\ref{scalefactorwm1vm})). For $m\geq 0$ the evolution of the two branes is very similar for a very small scale factor since $H_c^2\simeq H_0 ^2= constant$. In the case of $m<0$ the relation between the Hubble parameters is $H_c^2\simeq \frac{[1-m]^2}{[1-2m]^2} H_0 ^2$ then for more negative values of $m$ the expansion is slower for the hidden brane (see Fig. (\ref{hubblewm1})). The Hubble parameter diverges at the upper bound of the scale factor $a_0$ which correspond in this case to $m=1,2,3$. For late time cosmology and $m\leq 0$ the evolution of the two branes is very similar because $H_c ^2\simeq H_0 ^2=1/r_0 ^2$ (see Fig. (\ref{hubblewm1})). 

\begin{figure}[htp!]
  \includegraphics[width=0.45\textwidth]{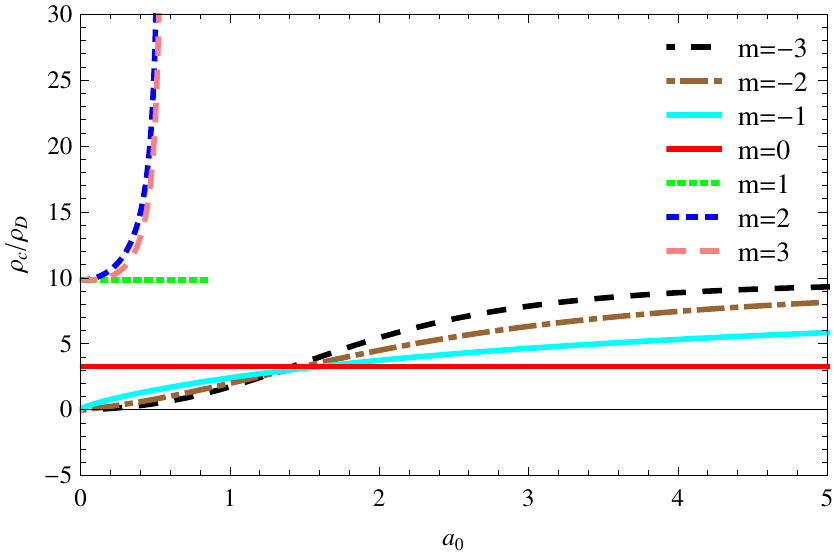}
  \includegraphics[width=0.45\textwidth]{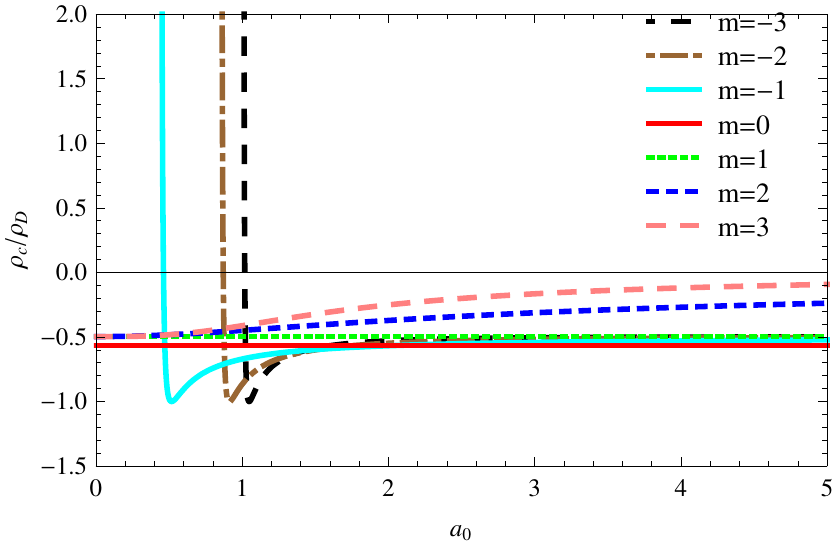}
  \caption{(left) The evolution of the energy density $\rho_c$ of the hidden brane as function of the scale factor of the visible brane $a_0$ with $w_0=-1$ for the self-accelerating branch. (right) The same as before but for the normal branch.}
  \label{rhocwm1}
\end{figure}

The behavior of the energy density is presented in Fig. (\ref{rhocwm1}), where it is shown that for $m<0$ it approaches to zero for $a\rightarrow 0$. For $m=1$ the energy density has the following constant value $\rho_0 =4\rho_D r_0 H_0[1+ r_0H_0]$ and for $m\geq 0$ it takes the same value for $a_0 \rightarrow 0$. In the low energy limit the energy density takes the following asymptotic values: for $m>1/2$ there is an upper bound for $a_c$ and the energy density diverges at this point, for $0<m<1/2$ its value approaches to zero and for $m<0$ it tends to $\rho_c = 4r_0\rho_D H_0 [1+r_0 H_0]$.

\begin{figure}[htp!]
  \includegraphics[width=0.45\textwidth]{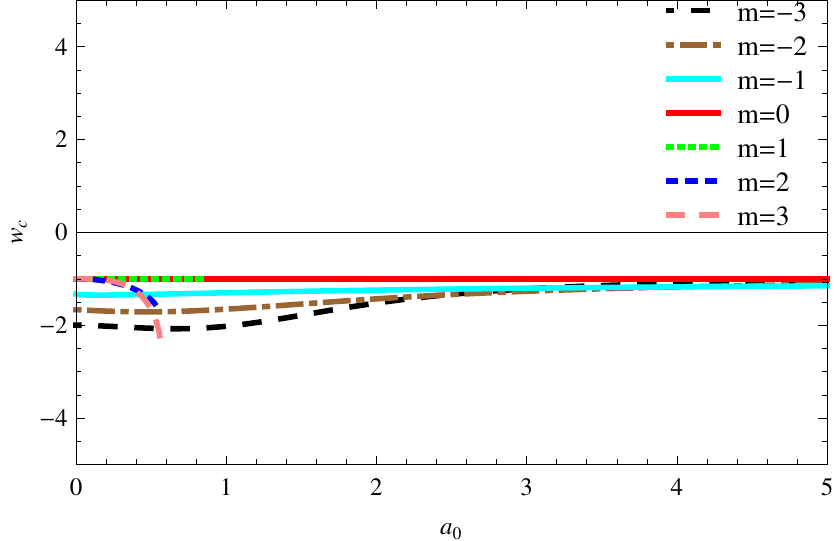}
  \includegraphics[width=0.45\textwidth]{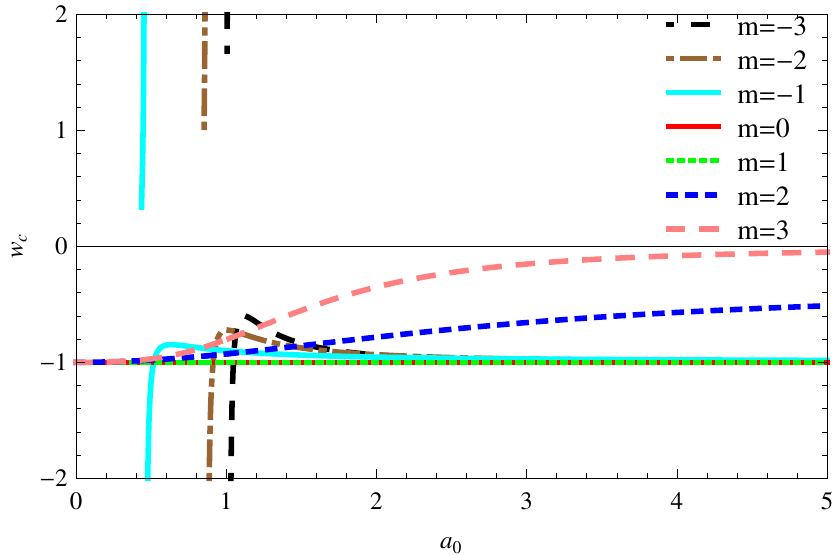}
  \caption{(left) The evolution of the equation of state parameter $w_c$ of the hidden brane as function of the scale factor of the visible brane $a_0$ with $w_0=-1$ for the self-accelerating branch. (right) The same as before but for the normal branch. We only plot $w_c$ in the allowed range for the scale factor $a_0$.}
  \label{wcparameterwm1}
\end{figure}

The equation of state parameter in the high energy limit when $a\rightarrow 0$ takes the asymptotic values of $w_c \rightarrow -1$ for $m>0$ and $w_c \rightarrow -1 + m/3$ for $m<0$. In the low energy limit, for very large value of the scale factor, the asymptotic form has the following expressions: $w_c \approx -1$ for $m<0$ and $w_c \approx -1 +m/3$ for $0<m<1/2$. In the case $m>1/2$ there is a upper bound for the scale factor $a_0$ and the equation of state parameter takes the value $w_c=-(2+m)/3$ for $1/2<m<1$ and $w_c = -(1 +2m)/3$  for $m>1$. The case $m=1$ produces a constant value for $w_c = -1$ for the allowed values of $a_c$ (See Fig. (\ref{wcparameterwm1})).

In the normal branch there is a bound from below for $a_0$ when $m<0$ (see Fig. (\ref{scalefactorwm1vm})). In the case $0<m<1/2$ there is a maximum value for the scale factor $a_0$ and for $m>1/2$ there are not restrictions for the scale factors of the two branes. The expansion of the hidden brane is slower for increasing positive values of $m$. The evolution of the two branes is very similar at the beginning of the universes for $m\geq 0$ since for very small scale factor $a_0$ we have $H_c^2\simeq H_0^2$ (see Fig. (\ref{hubblewm1})). For $m<0$ the Hubble parameter diverges at the minimum value of the scale factor $a_0$. In late time cosmology we have that for $m\geq 0$ and $m\neq 1$ the Hubble parameters are related by $H_c^2\simeq \frac{[1-m]^2}{[1-2m]^2} H_0 ^2$. For $m<0$ the evolution of the two branes is very similar since $H_c^2\simeq H_0 ^2$ (see Fig. (\ref{hubblewm1})) and for $m=1$ the expansion of the hidden brane stops since $H_c \simeq 0$ for very large values of the scale factor $a_0$. 

 For $m<0$ the energy density diverges when the scale factor $a_0$ takes its minimum value (see Fig. (\ref{rhocwm1})). For $m=1$ the energy density is a constant $\rho_c = \rho_D 4r_0H_0[r_0H_0 -1]$ which coincide with its asymptotic value for $a_0 \rightarrow 0$ for the cases of positive $m$. In the low energy limit for $m<0$ we have that the following constant value $\rho_0 =4r_0H_0[r_0H_0-1]$ for very large values of the scale factor $a_0$. In the case of $0<m<1/2$ there is a bound from above for $a_0$ and in this point the energy density diverges. For $1/2<m<1$ the asymptotic value of the energy density diverges and for $m>1$ the energy density tends to zero.
 
The equation of state parameter diverges for $m=-1,-2,-3$ due to the energy density is zero at some value of the scale factor $a_0$ (See Fig. (\ref{wcparameterwm1})). Besides, in the high energy limit the asymptotic value of the equation of state parameter takes the value $w_c \approx -1$ for $a\rightarrow 0$ and $m>0$. For $m<0$ there is bound from below $a_b$ and in this point $w_c = -(1+2m)/3$. In the low energy limit, when $m<0$ we have $w_c \rightarrow -1$ for $a_0 \rightarrow \infty$. For $0<m<1/2$ there is a bound from above and in this point $w_c = -(2m+1)/3$. For $m>1/2$ there are not restrictions and we have the following asymptotic limits: $w_c = -1 +2m/3$ for $1/2<m<1$ and $w_c=-1 +m/3$ for $m>1$.

\section{Concluding remarks}
The idea that our Universe could be a hypersurface embedded in a higher dimensional spacetime has attracted a lot of interest for a long time. In particular, in models inspired by string theory, a possible viable cosmological model consists of two branes embedded in a 5-dimensional background with a compactified extra dimension.

In this paper we have investigated the cosmology of two 3-branes with $Z_2$ symmetry embedded in a 5-dimensional Minkowski spacetime with a compactified extra dimension in the case when an intrinsic curvature term is included in the action for each brane. By means of the 5-dimensional Einstein equations and the junction conditions we have found global solutions for the scale factor of the hidden brane in terms of the scale factor of the visible brane  which give as a result possible restrictions on the allowed values of the scale factor of the visible brane. 

Additionally, the global solutions for the other metric functions impose topological restrictions between the energy densities, the Hubble parameters and the equation of state parameters of the two branes

For example, a very interesting feature of this kind of models is the existence of a relation between the energies on the two branes due to the compactified extra dimension. This relation implies, once we fix the energy content in the visible brane, a specific form for the energy and pressure of the hidden brane in terms of its corresponding quantities in the other brane. 

The local solutions reproduce the well-known behavior of DGP cosmology when there exists a crossover length $r_0$ between the 4 and 5-dimensional gravity regimes.  

The cosmology in the visible brane has two branches (the self-accelerating and normal) related with the sign of the total energy ($-\epsilon$) which include the perfect fluid and the effective part calculated from the intrinsic curvature of the brane.

For the self-accelerating branch where $\epsilon=1$, the global solutions impose restrictions for the allowed values of the scale factor $a_0$ of the visible brane.  The possible cosmologies for the visible brane include an oscillating universe and an upper bound for the scale factor or recollapsing universe.

For the normal branch where $\epsilon=-1$, the cosmologies include a universe without initial singularity or Big Bounce cosmology and an upper bound for the scale factor or recollapsing universe. The oscillating scenario is not present in this branch. The case of a lower bound for the scale factor of the visible brane is very interesting since the minimum scale factor could be as the same order of the scale factor of the Planck epoch where the initial singularity is avoided. In the other case, the maximum scale factor could be very large and still this could lead to a viable cosmology model.

The topological restriction between the total energy densities  produces a relation between the branches of the two DGP branes. If the visible brane is described by means of the self-accelerating (normal) branch then the hidden brane brane have to be described by the normal (self-accelerating) branch.

Besides, we have obtained the high and low energy limits for the cosmology of the two branes and analyzed its behaviors. We have studied the energy density, the Hubble parameter and the equation of state parameter of the hidden brane in terms of the scale factor of the visible brane when we consider only one component of the perfect fluid in this brane.

Finally, it is very interesting to investigate the evolution of the two branes when it is present a scalar field with a potential which could model the inflationary epoch in the visible brane.

\section*{Acknowledgments} 
This work was partially supported by SNI-M\'exico, COFAA-IPN, EDI-IPN, SIP-IPN: 20180741, 20195330, 20201841 number projects.

\end{document}